\documentclass[a4paper]{jpconf}
\usepackage{graphicx}
\usepackage[T2A]{fontenc}
\usepackage[utf8]{inputenc}
\usepackage[english]{babel}

\begin{document}

\title{SANC system and its applications for LHC}

\author{R Sadykov, A Arbuzov, D Bardin, S Bondarenko, P Christova,\\
L Kalinovskaya, V Kolesnikov, A Sapronov and E Uglov}

\address{Joint Institute for Nuclear Research, Joliot-Curie str. 6, Dubna, 141980, Russia}

\ead{renat.sadykov@cern.ch}

\begin{abstract}
The {\tt SANC} computer system is aimed at support of analytic and numeric
calculations for experiments at colliders. The system is reviewed briefly.
Recent results on high-precision description of the Drell-Yan processes at
the LHC are presented. Special attention is paid to the evaluation of
higher order final-state QED corrections to the single $W$
and $Z$ boson production processes. A new Monte Carlo integrator {\tt mcsanc}
suited for description of a series of high-energy physics processes at the
one-loop precision level is presented.
\end{abstract}

\section{Introduction}

After the discovery at the LHC of a new particle which is very likely the long
awaited Higgs boson, the precision tests of the Standard Model (SM)
became of crucial importance. Moreover, the absence of clear signals of
new physics at the LHC stimulates comparisons of high-precision predictions 
received within the SM with the experimental data.
The rather accurate measurement of the Higgs boson mass
performed at the LHC fixed the last free parameter of the SM.
This potentially allows to compute theoretical predictions by means
of perturbation theory to a very high precision level. The corresponding
experimental accuracy is continuously growing up with accumulating of statistics
and development of analysis techniques. To provide adequately accurate theoretical
predictions, one has to take into account and combine different relevant effects.

\section{SANC Project}

{\tt SANC} is a computer system for Support of Analytic and Numeric
calculations for experiments at Colliders~\cite{Andonov:2004hi,Bardin:2005dp}.
It can be accessed through the Internet at
{\verb"http://sanc.jinr.ru/"}.
The {\tt SANC} system is suited for calculations of NLO QED, electroweak (EW), and QCD
radiative corrections (RC) to various SM processes.
Automatized analytic calculations in {\tt SANC} provide
{\tt FORM} and {\tt FORTRAN} modules~\cite{Andonov:2008ga}.
Stand-alone {\tt FORTRAN} modules
for differential cross sections as well as Monte Carlo programs
(integrators and generators) are available to download.

\begin{figure}[ht]
\begin{center}
\includegraphics[width=8cm]{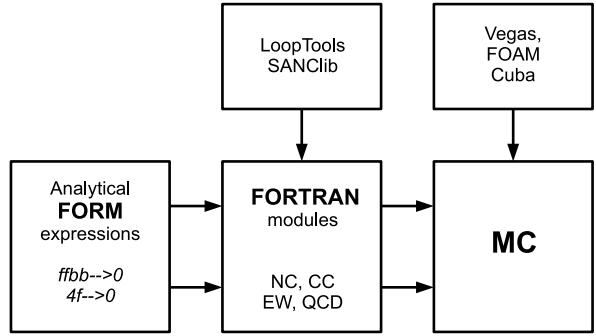}
\end{center}
\caption{\label{SANC_flow} A schematic flowchart of the {\tt SANC} system.}
\end{figure}

A schematic flowchart of the {\tt SANC} system is presented in Fig.~\ref{SANC_flow}.
In the core of the system, analytic calculations of amplitudes, cross sections,
and other quantities for various processes are computed by means of the {\tt FORM}
language~\cite{Kuipers:2012rf}. The results of analytic calculations are translated
automatically into {\tt FORTRAN} modules which can be used by external programs.
The modules extensively use internal and external~\cite{Hahn:2006qw}
{\tt FORTRAN} libraries of Passarino-Veltman functions.

\section{Description of Drell-Yan processes in SANC}

High-precision theoretical description of Drell-Yan (DY) like processes
(i.e. single $Z$ and $W$ boson production) at the LHC
is of great importance. In fact, these processes
have large cross sections, clear signatures in detectors, and they are rather
well suited for tests of the SM and searches for new physics.
So DY processes
provide {\em standard candles} for detector calibration at the LHC and the Tevatron.
Single $Z$ and $W$ boson production is also used at the LHC for extraction
of parton distribution functions (PDF) in the kinematical region which has not
been accessed by earlier experiments. It is planned to get the most precise
experimental values for the mass and width of $W$ boson
using the future high statistics data on the charged current DY process.
DY processes will also provide background to many other reactions being of interest at the LHC.
Moreover, some new physics searches like the ones for contact four-fermion interactions
will be performed in these channels.
Therefore it is crucial to control the theoretical predictions for production
cross sections and kinematic distributions of both the neutral current (NC)
and charged current (CC) DY processes.

The experimental precision tag for inclusive observables in DY processes
is about 1\%. That means we need to provide the accuracy of theoretical predictions
of about 0.3\% in order not to spoil the results of the LHC data analysis.
This is a challenge for the theory.
Aiming at high precision of DY description we need to take into account the following effects:
\begin{itemize}

\item
QCD contributions at LO, NLO and NNLO;

\item
parton showers and hadronization effects;

\item
EW RC at one-loop at least;

\item
most important higher order EW contributions (re-summed where possible);

\item
an interplay of QCD and EW corrections;

\item a tuned input: coupling constants, the hadronic vacuum polarization, and
PDF for the appropriate energy scales and $x$-values.

\end{itemize}

All relevant effects should be implemented in a Monte Carlo event generator which
can be directly used in the experimental data analysis.
Actually, treating all the listed effects in a single code is a very involved
task. More realistic is the possibility to make a chain of event generators, which pass
generated events one to another and {\em dress} them with additional effects.

\begin{table}[ht]
\caption{\label{DY_comparisons} Tuned comparisons between {\tt SANC} and other codes
in predictions on DY processes.}
\begin{center}
\begin{tabular}{ccc}
\hline
Processes & Codes & Ref.\\
\hline
&&\\
CC DY (NLO EW)
& { \tt SANC}, { \tt HORACE}, { \tt WGRAD2}
& {\cite{Gerber:2007xk}}\\
&&\\
\hline
&&\\
NC DY (NLO EW)
& { \tt SANC}, { \tt HORACE}, { \tt ZGRAD2}
& {\cite{Buttar:2008jx}}\\
&&\\
\hline
&&\\
CC \& NC DY
& {\tt SANC}, { \tt HORACE}, { \tt WZGRAD}, { \tt RADY},
& \cite{ew-lpcc:www}\\
(NLO QCD \& EW)
& { \tt FEWZ}, { \tt DYNNLO}, { \tt POWHEG-w}, { \tt POWHEG-z} &\\
&&\\
\hline
\end{tabular}
\end{center}
\end{table}

\begin{figure}[ht]
\begin{center}
\includegraphics[width = 0.45\textwidth]{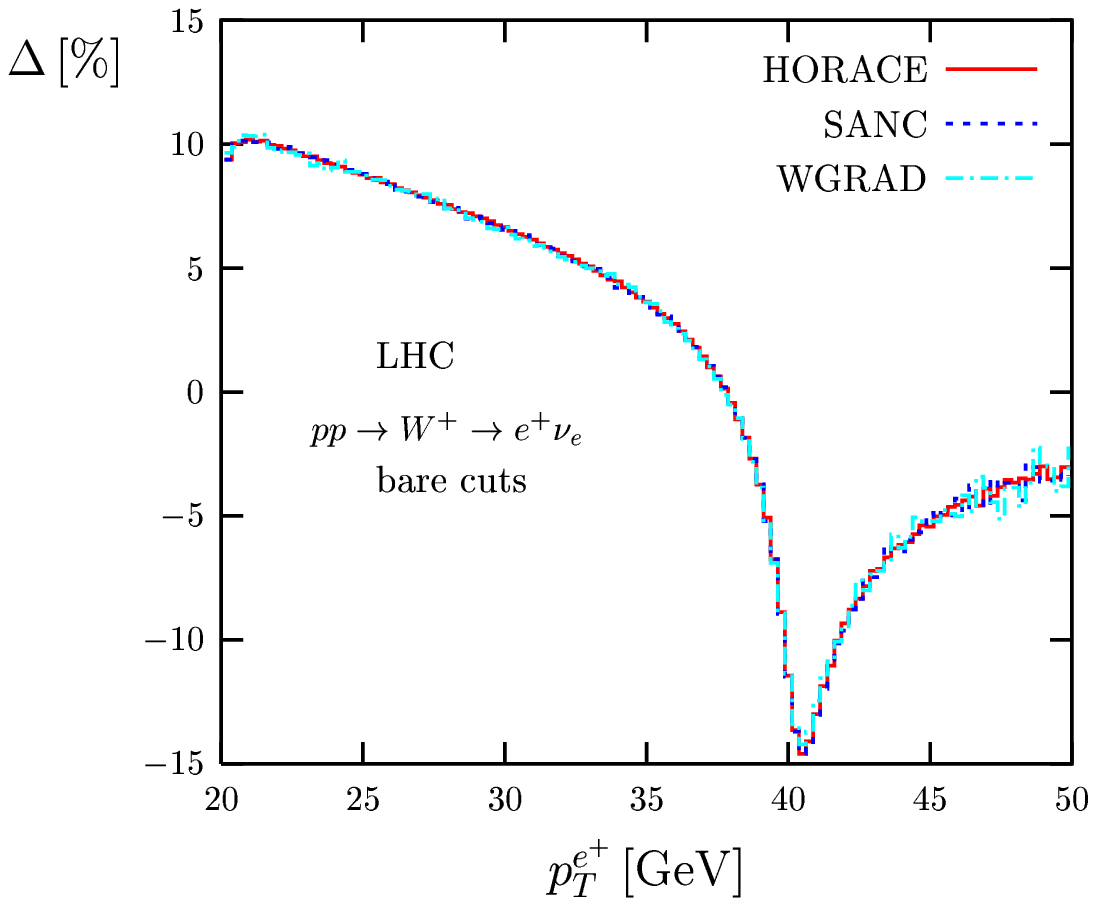}
\includegraphics[width = 0.45\textwidth]{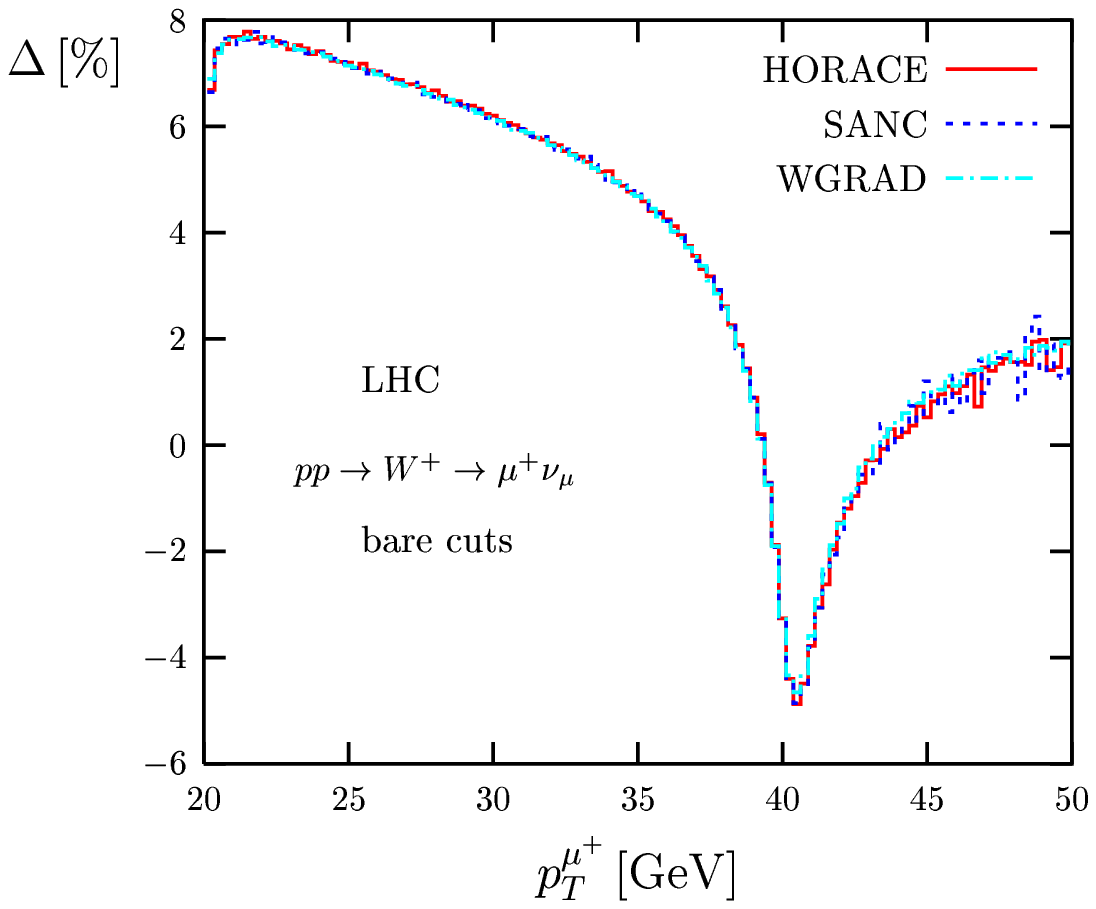}\\
\end{center}
\caption{\label{fig_PT_CC} Tuned comparison of positron (left) and anti-muon (right)
transverse momentum distribution in the CC DY process~\cite{Gerber:2007xk}.}
\end{figure}

\begin{figure}[ht]
\begin{center}
\includegraphics[width = 0.45\textwidth]{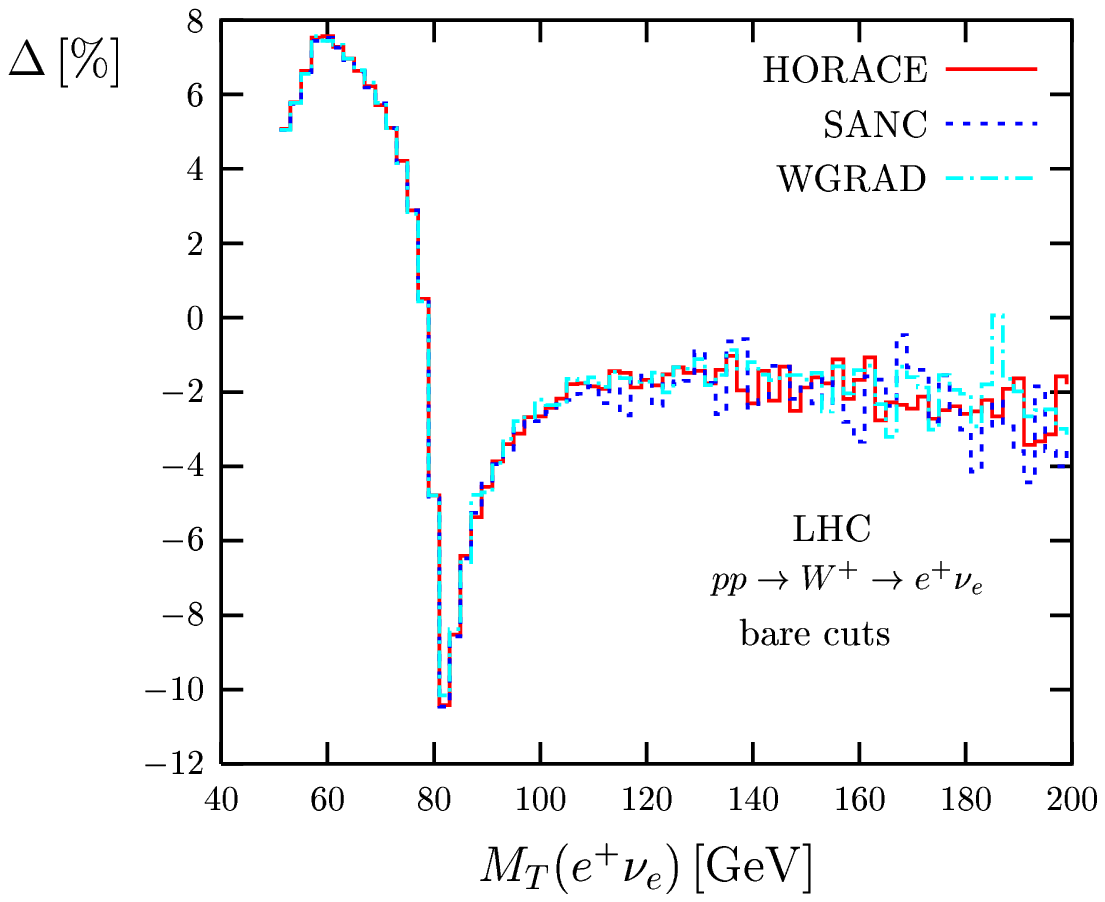}
\includegraphics[width = 0.45\textwidth]{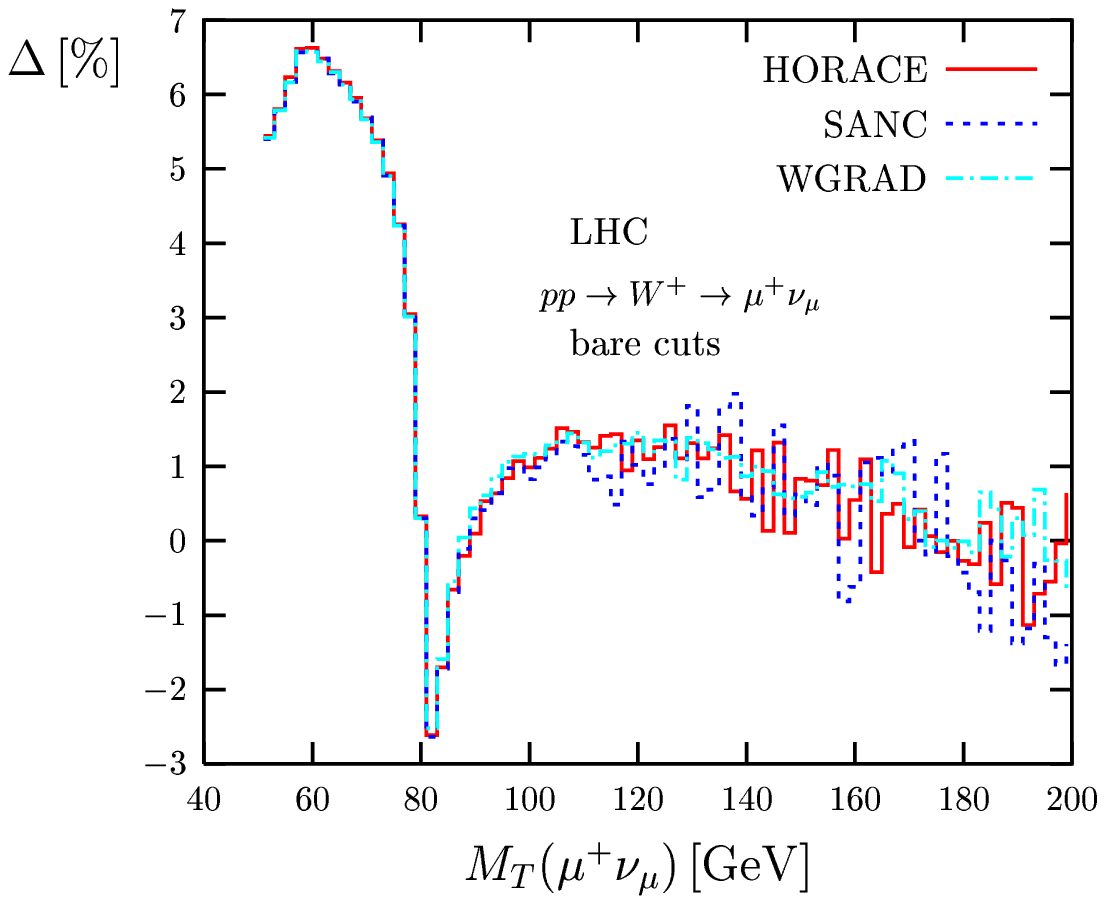}
\end{center}
\caption{\label{fig_MT_CC} Tuned comparison of $e^+\nu_e$ (left) and $\mu^+\nu_\mu$ (right)
transverse mass distribution in the CC DY process~\cite{Gerber:2007xk}.}
\end{figure}

\begin{figure}[ht]
\begin{center}
\includegraphics[width = 0.45\textwidth]{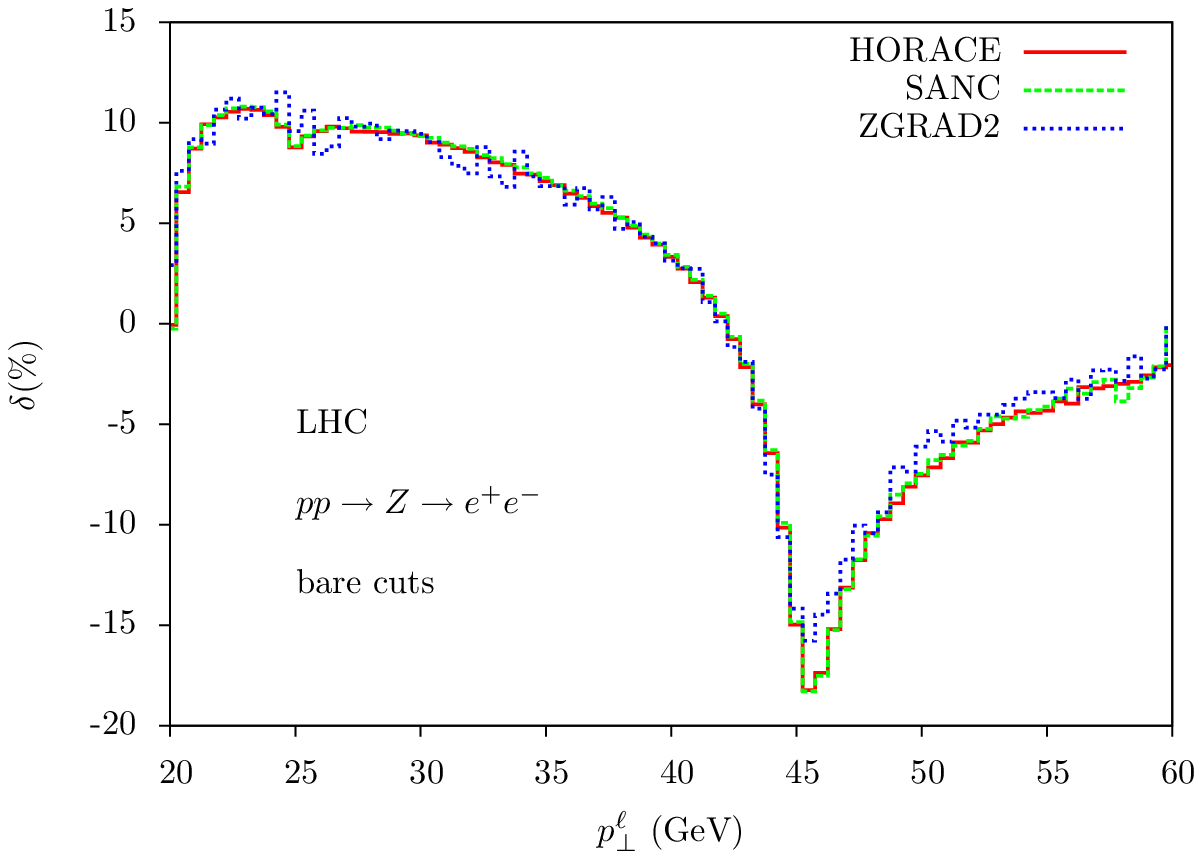}
\includegraphics[width = 0.45\textwidth]{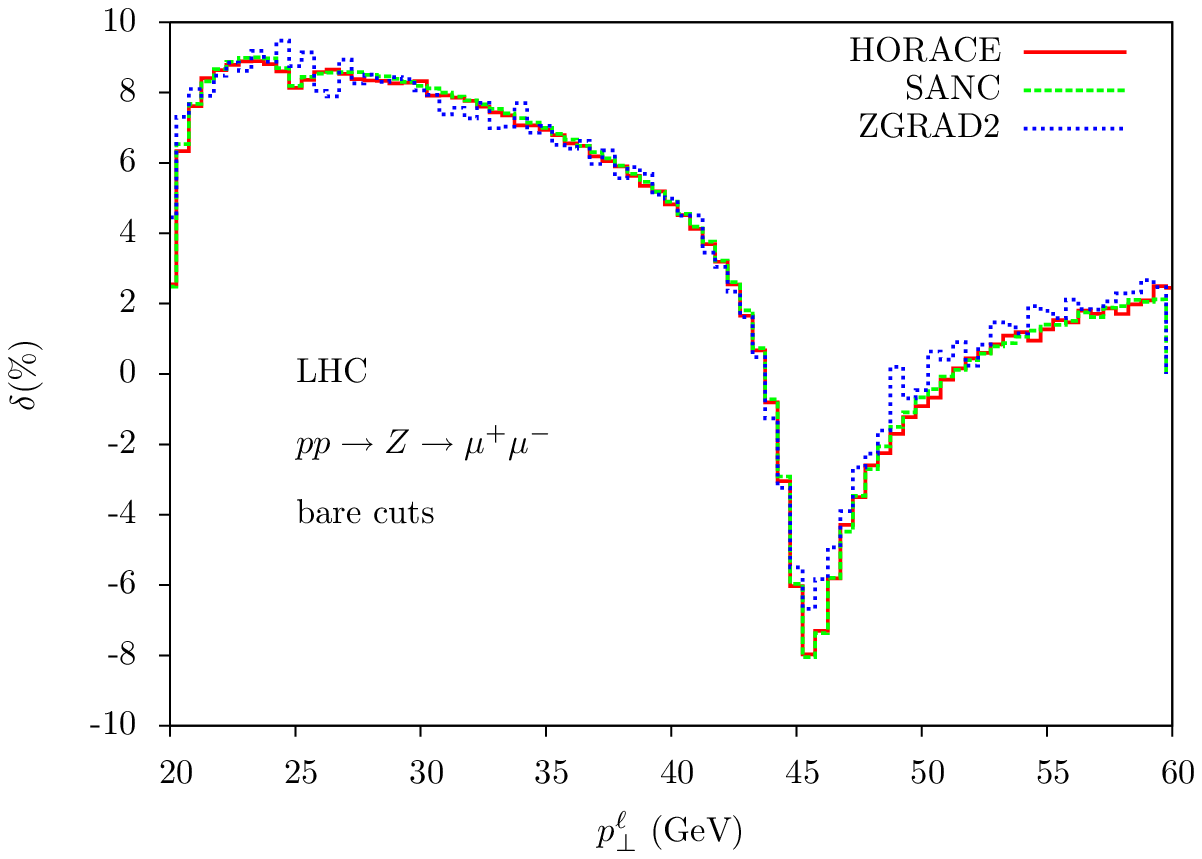}
\end{center}
\caption{\label{fig_PT_NC} Tuned comparison of positron (left) and anti-muon (right)
transverse momentum distribution in the NC DY process~\cite{Buttar:2008jx}.}
\end{figure}

\begin{figure}[ht]
\begin{center}
\includegraphics[width = 0.45\textwidth]{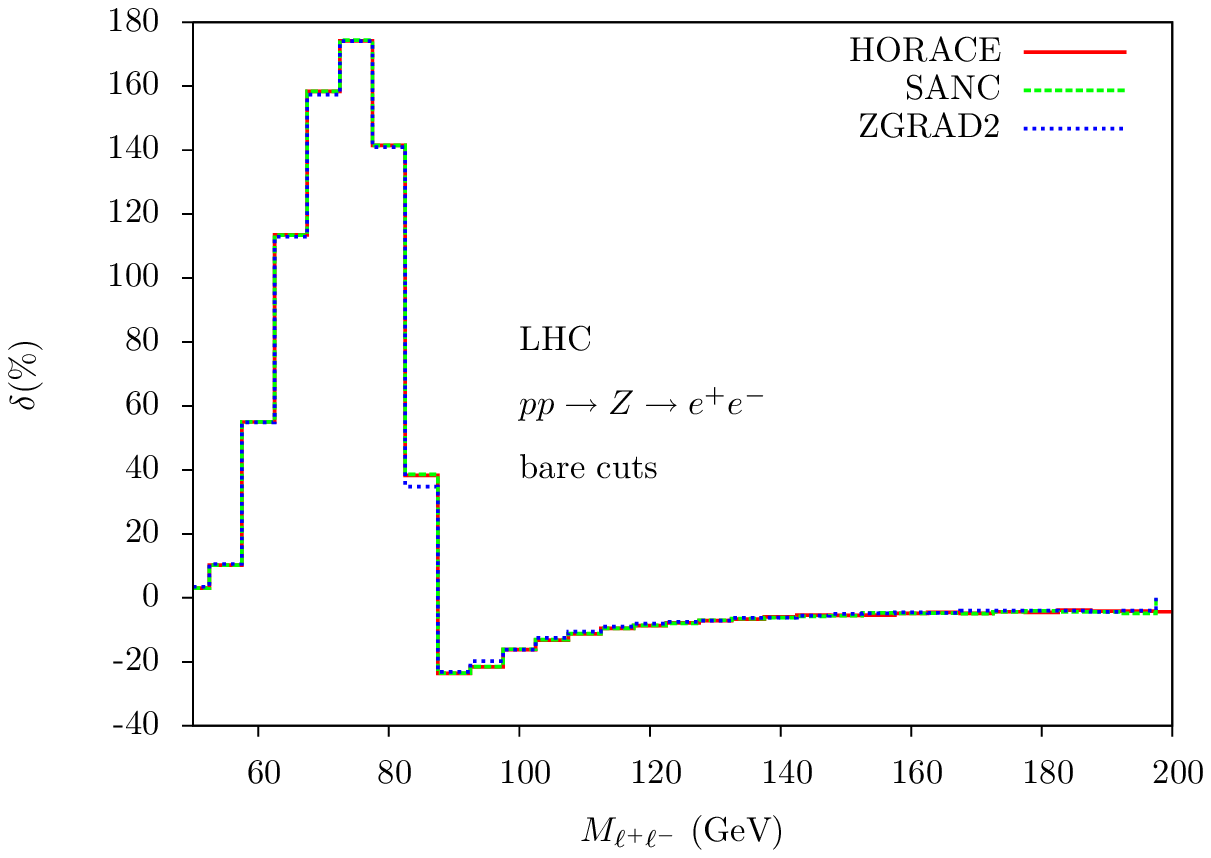}
\includegraphics[width = 0.45\textwidth]{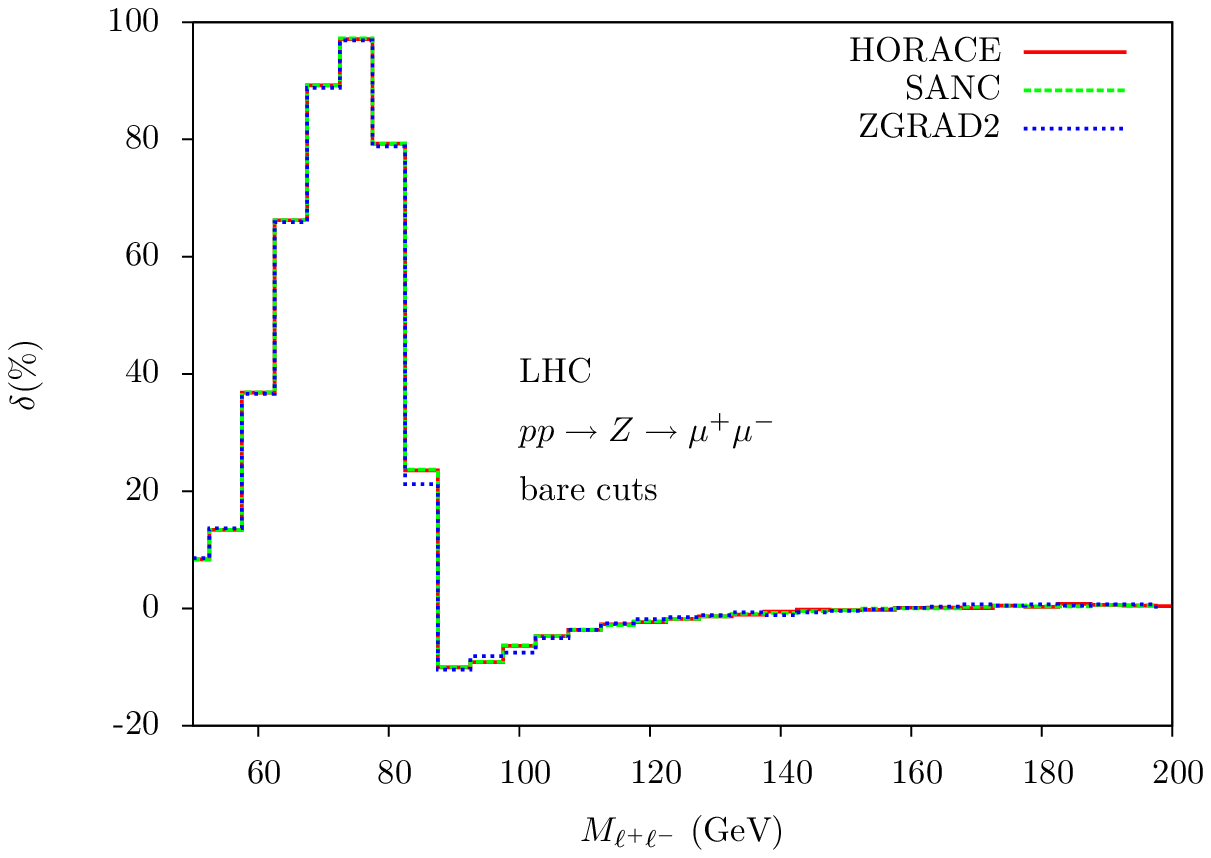}
\end{center}
\caption{\label{fig_M_NC} Tuned comparison of $e^+e^-$ (left) and $\mu^+\mu^-$ (right)
invariant mass distribution in the NC DY process~\cite{Buttar:2008jx}.}
\end{figure}

With a tuned set-up we got an excellent agreement with several other groups
in the values of EW RC to the CC~\cite{Gerber:2007xk}
and NC~\cite{Buttar:2008jx} DY processes, see Table~\ref{DY_comparisons}.
Examples of differential distributions of the EW RC contributions are given
in Fig.~\ref{fig_PT_CC} and Fig.~\ref{fig_MT_CC} for the CC channel and in
Fig.~\ref{fig_PT_NC} and Fig.~\ref{fig_M_NC} for the NC one.
The relative effect of EW RC is defined as
\begin{eqnarray}
\Delta = \frac{\sigma(\mathcal{O}(\alpha))-\sigma(\mathrm{Born})}{\sigma(\mathrm{Born})}\cdot 100\%.
\end{eqnarray}
A new series of tuned comparisons is going on within the $W$-mass workshop~\cite{ew-lpcc:www}.
QCD corrections are also taken into consideration, and the set-up is
adjusted to simulate the experimental conditions.

\section{Final State Radiation in single $W$ and $Z$ production}

Numerically, the effect of QED final-state radiation (FSR) off charged leptons
at the LHC is very large. The corresponding correction can reach, {\it e.g.}, even about 200\%
in the distribution of $e^+e^-$ invariant mass in NC DY.
For this reason FSR deserves special attention.
Recently a dedicated study~\cite{Arbuzov:2012dx} of the effect was performed by the joint effort
of {\tt SANC} and {\tt PHOTOS}~\cite{Davidson:2010ew} groups.

Our goals were:
\begin{itemize}
\item
to perform a comparison between {\tt SANC} and {\tt PHOTOS}
in the {\em single} and {\em multiple} photon modes for FSR in DY processes
both for neutral and charged currents;

\item
to check that QED FSR is properly installed in the programs;

\item
to tune separation of the FSR QED corrections from the complete EW NLO ones.
\end{itemize}

In {\tt PHOTOS} the bremsstrahlung corrections to decays of $W$ and
$Z$ bosons are treated in a special way in the next-to-leading approximation
with exponentiation. {\tt PHOTOS} is linked to the {\tt PYTHIA} program
in the standard Monte Carlo simulation through {\tt HepMC} interface.
In {\tt SANC} the complete EW corrections at one-loop are calculated for
single $W$ and $Z$ production. The FSR QED corrections can be
separated from the rest of EW corrections by means of flags in
{\tt SANC} Monte Carlo event generators and integrators.

For a tuned comparison of FSR effects in {\tt SANC} and {\tt PHOTOS} we used the
following set-up. Parton level cross-section was convoluted with CTEQ6L1 PDF
set with running scale $Q^2 = s$, where $s$ is the total energy in the center-of-mass system
of partons. The C++ version of {\tt PHOTOS} program was used
together with {\tt Pythia8} which provided the Born-level events in
{\tt HEPevt} format that subsequently was passed to {\tt PHOTOS} to simulate photon
radiation off the final charged leptons. {\tt PHOTOS} was running in single and multiple
photon modes with matrix-element corrections turned on.
Here the relative corrections are defined as:

\begin{equation}
\delta = \frac{\sigma(\mathcal{O}(\alpha) FSR) - \sigma(\mathrm{Born})}{\sigma(\mathrm{Born})}
\end{equation}
for single photon radiation, and

\begin{equation}
\delta_{h.o.} = \frac{\sigma(\mathrm{h.o. FSR}) - \sigma(\mathcal{O}(\alpha) FSR)}{\sigma(\mathrm{Born})}
\end{equation}
for higher order FSR contributions.

\begin{figure}[ht]
\begin{center}
\includegraphics[width = 0.45\textwidth]{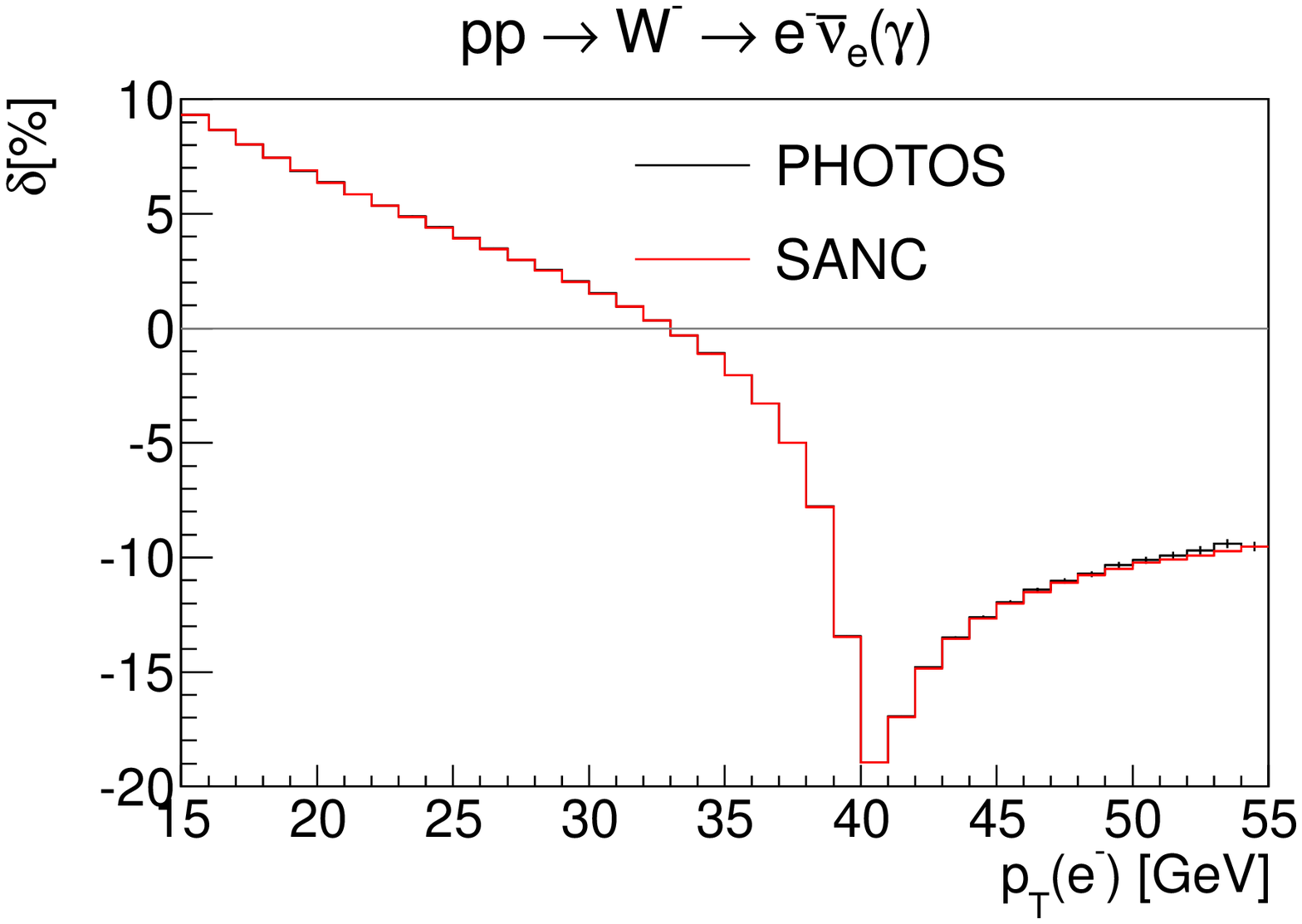}
\includegraphics[width = 0.45\textwidth]{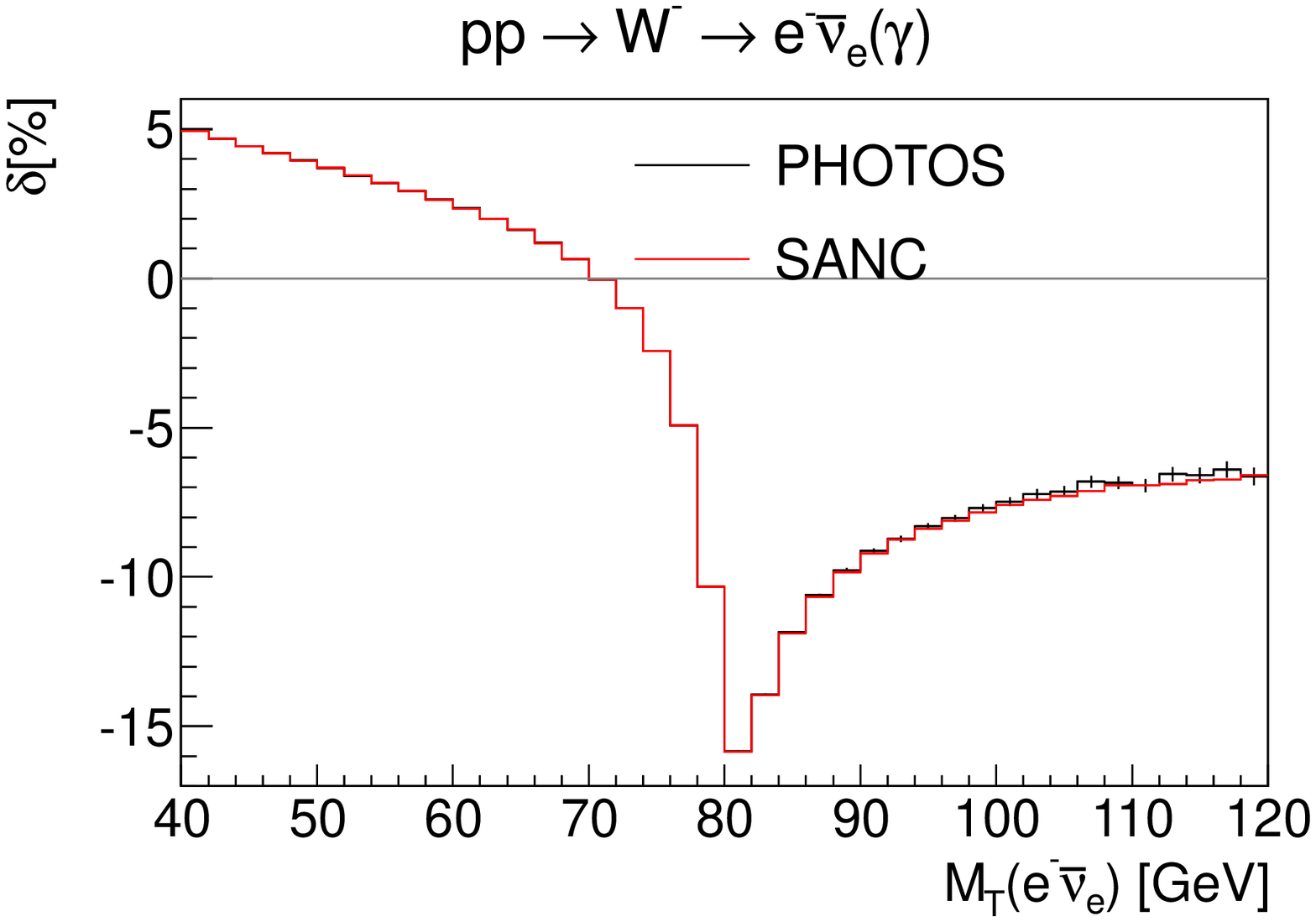}
\end{center}
\caption{\label{fig_sing_cc_bare} Tuned comparison of transverse momentum (left)
and transverse mass (right) distributions in CC DY.}
\end{figure}

\begin{figure}[ht]
\begin{center}
\includegraphics[width = 0.45\textwidth]{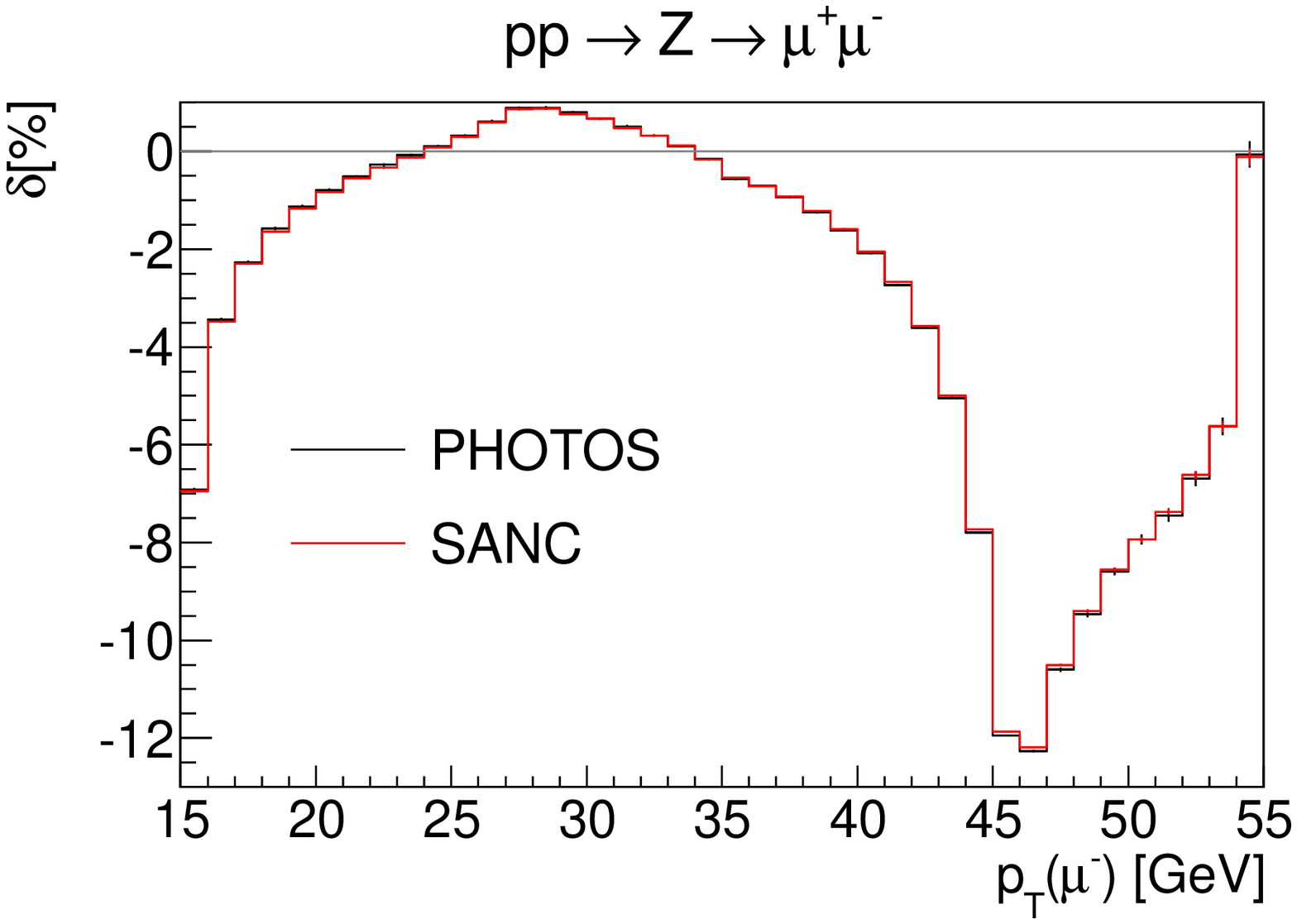}
\includegraphics[width = 0.45\textwidth]{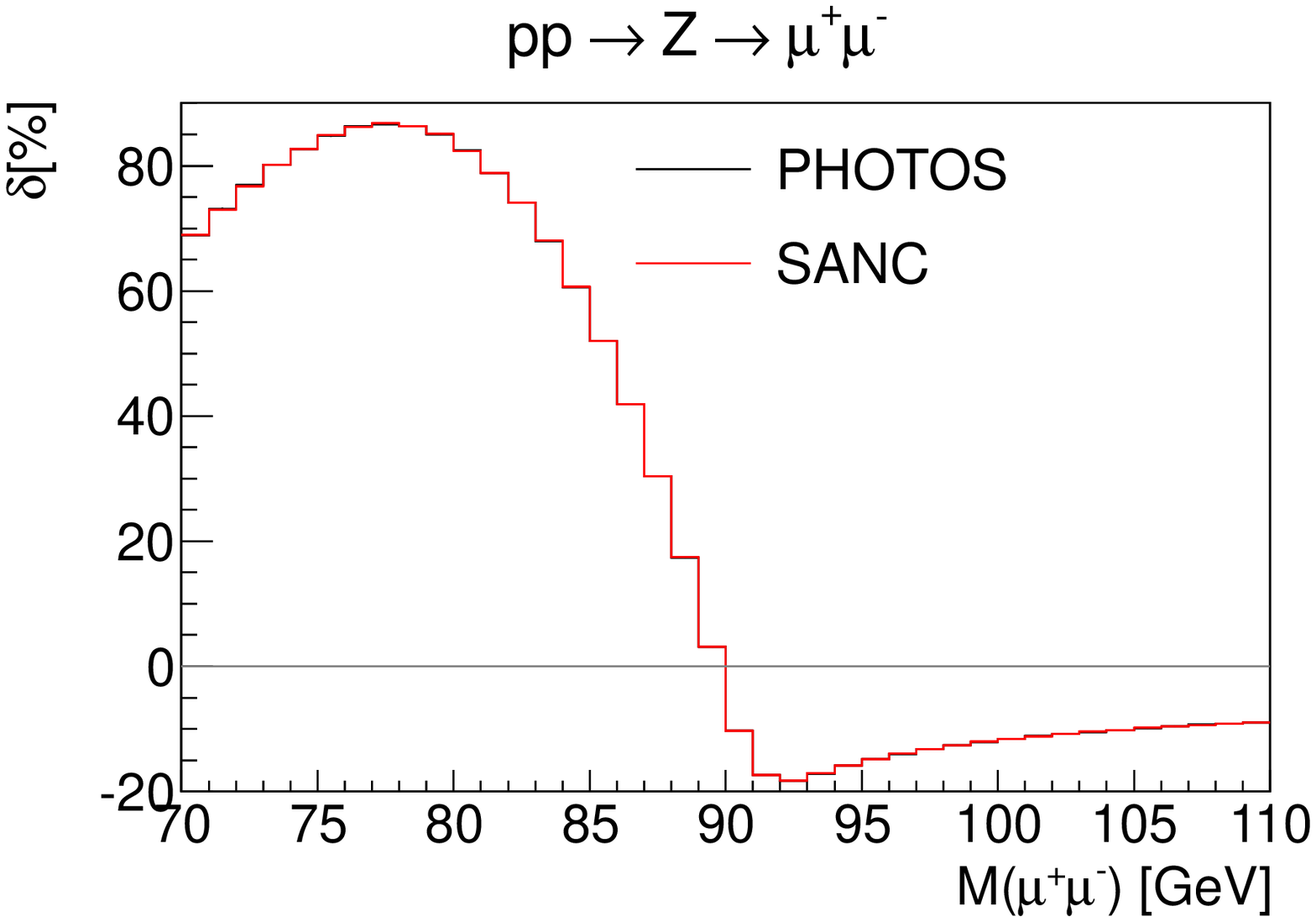}
\end{center}
\caption{\label{fig_sing_nc_pt_m} Tuned comparison of transverse momentum (left)
and invariant mass (right) distributions in NC DY.}
\end{figure}

\begin{figure}[ht]
\begin{center}
\includegraphics[width = 0.45\textwidth]{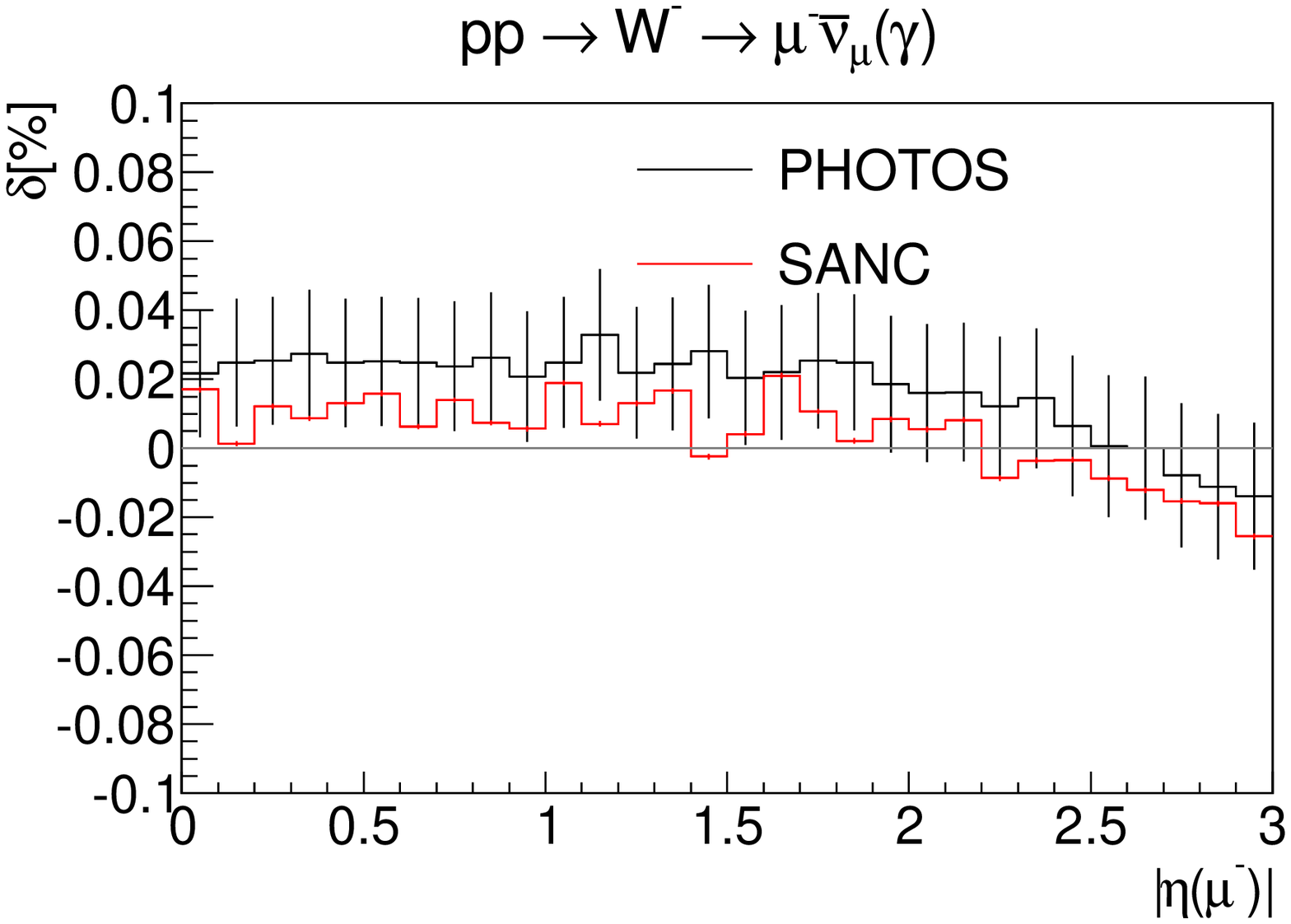}
\includegraphics[width = 0.45\textwidth]{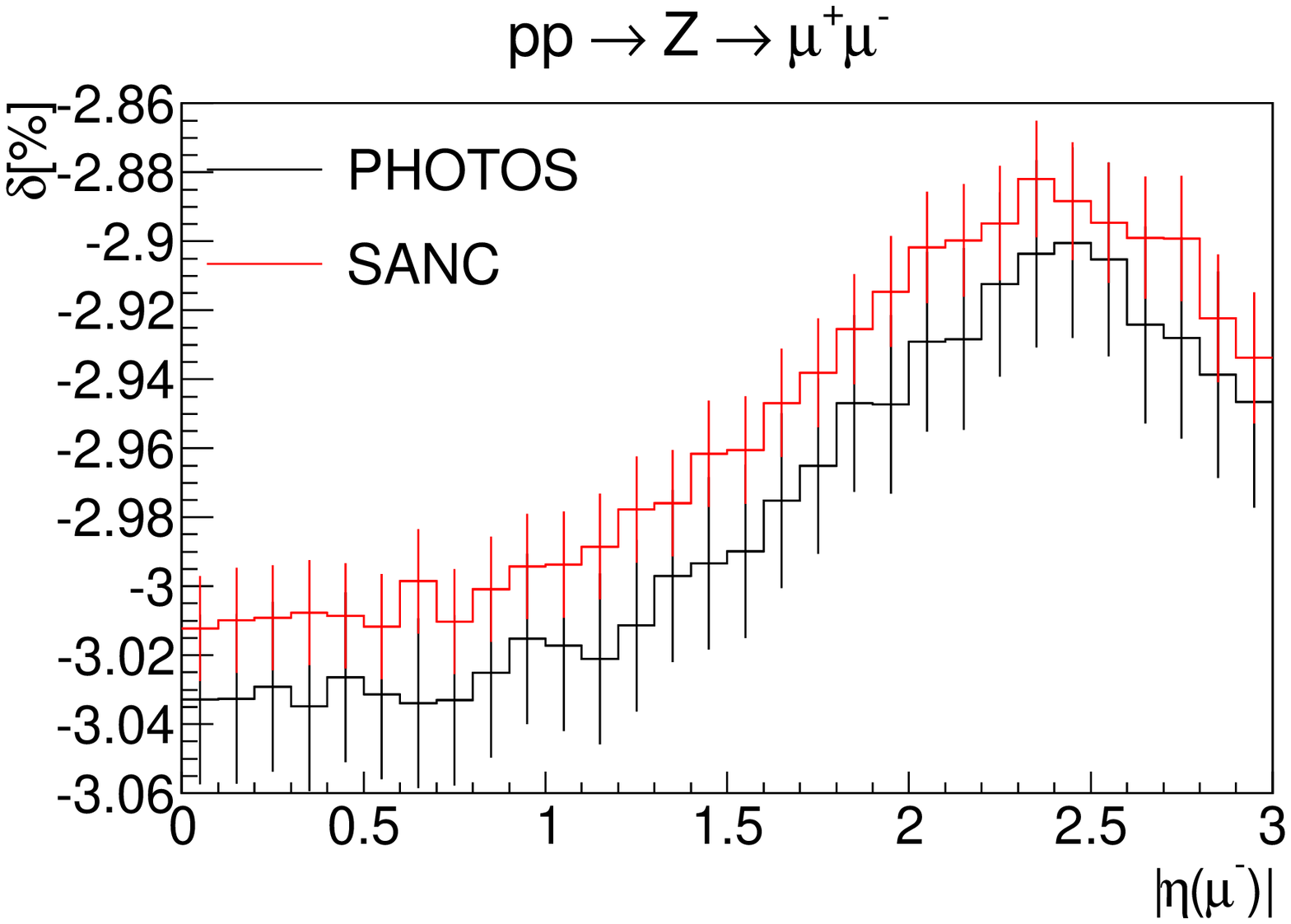}
\end{center}
\caption{\label{fig_sing_eta} Tuned comparison of lepton pseudorapidity
distribution in CC (left) and NC (right) DY processes.}
\end{figure}

Let us look at the comparison of a single photon, {\it i.e.} $\mathcal{O}(\alpha)$, FSR contribution
to CC and NC DY processes presented in Fig.~\ref{fig_sing_cc_bare} and Fig.~\ref{fig_sing_nc_pt_m},
respectively.
The distributions in the lepton transverse momentum and transverse (invariant) mass agree pretty well.
Nevertheless, the implementations of the FSR effect in {\tt SANC} and {\tt PHOTOS}
are different: there is a small systematic shift, as can be seen from the corresponding
distribution in the lepton pseudorapidity, see Fig.~\ref{fig_sing_eta}.
This shift is due to the specific feature of the {\tt PHOTOS} program: simulation of
FSR by means of {\tt PHOTOS} doesn't change the total cross section, while
the true $\mathrm{O}(\alpha)$ FSR makes a small change.

\begin{figure}[ht]
\begin{center}
\includegraphics[width = 0.45\textwidth]{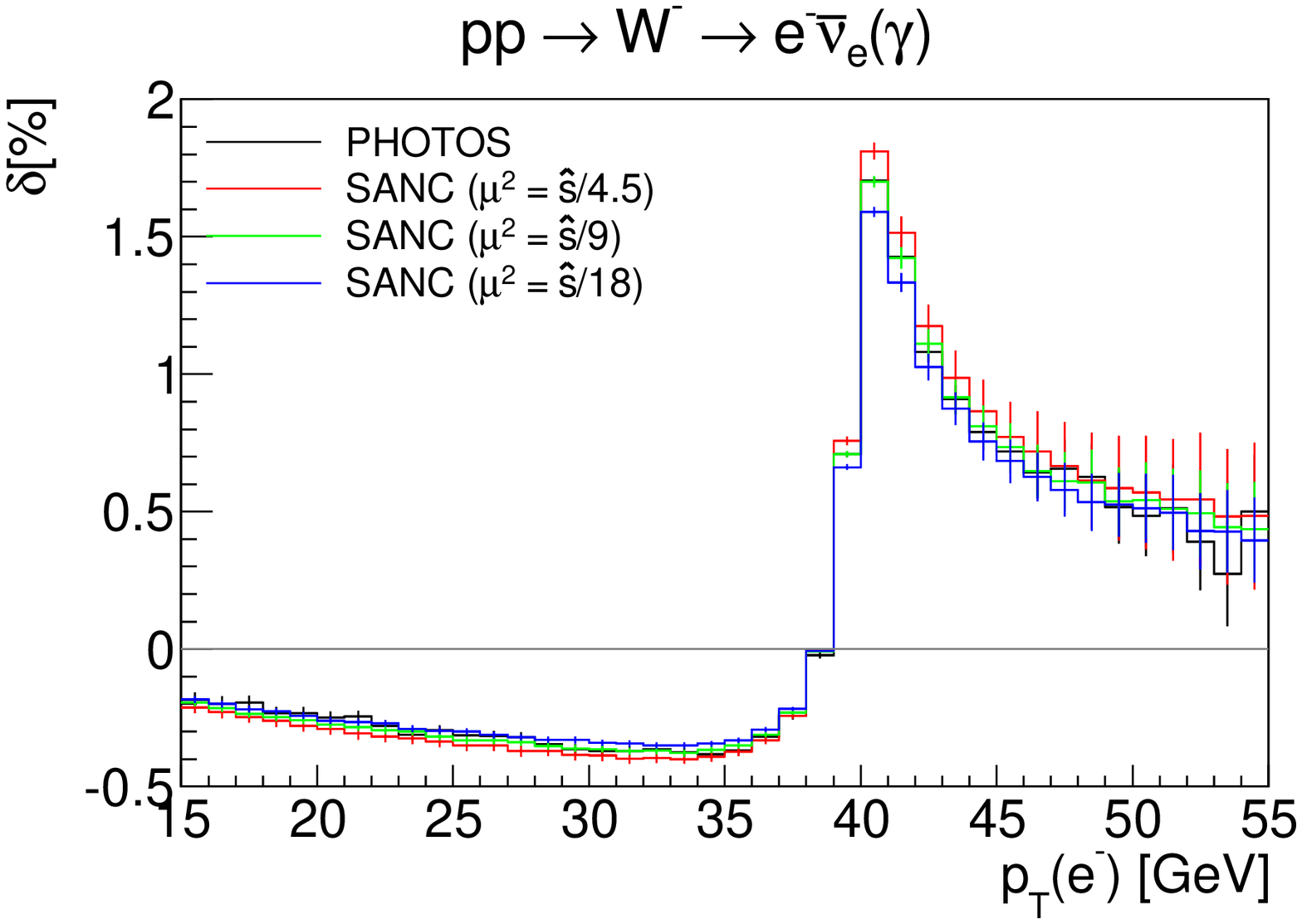}
\includegraphics[width = 0.45\textwidth]{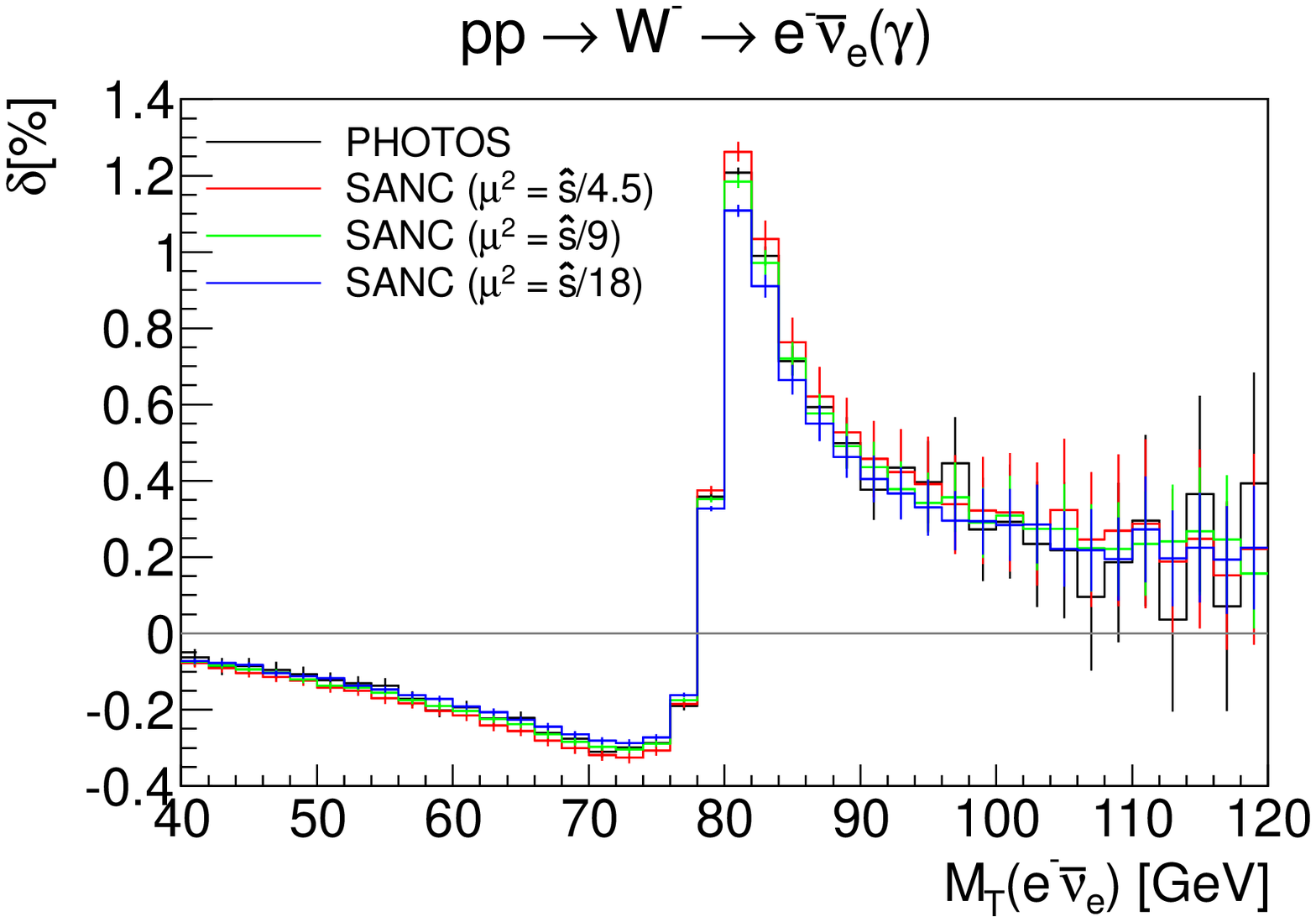}
\end{center}
\caption{\label{fig_multi_e} Comparison of higher order FSR contribution
to transverse momentum (left) and transverse mass (right) distributions in CC DY.}
\end{figure}

\begin{figure}[ht]
\begin{center}
\includegraphics[width = 0.45\textwidth]{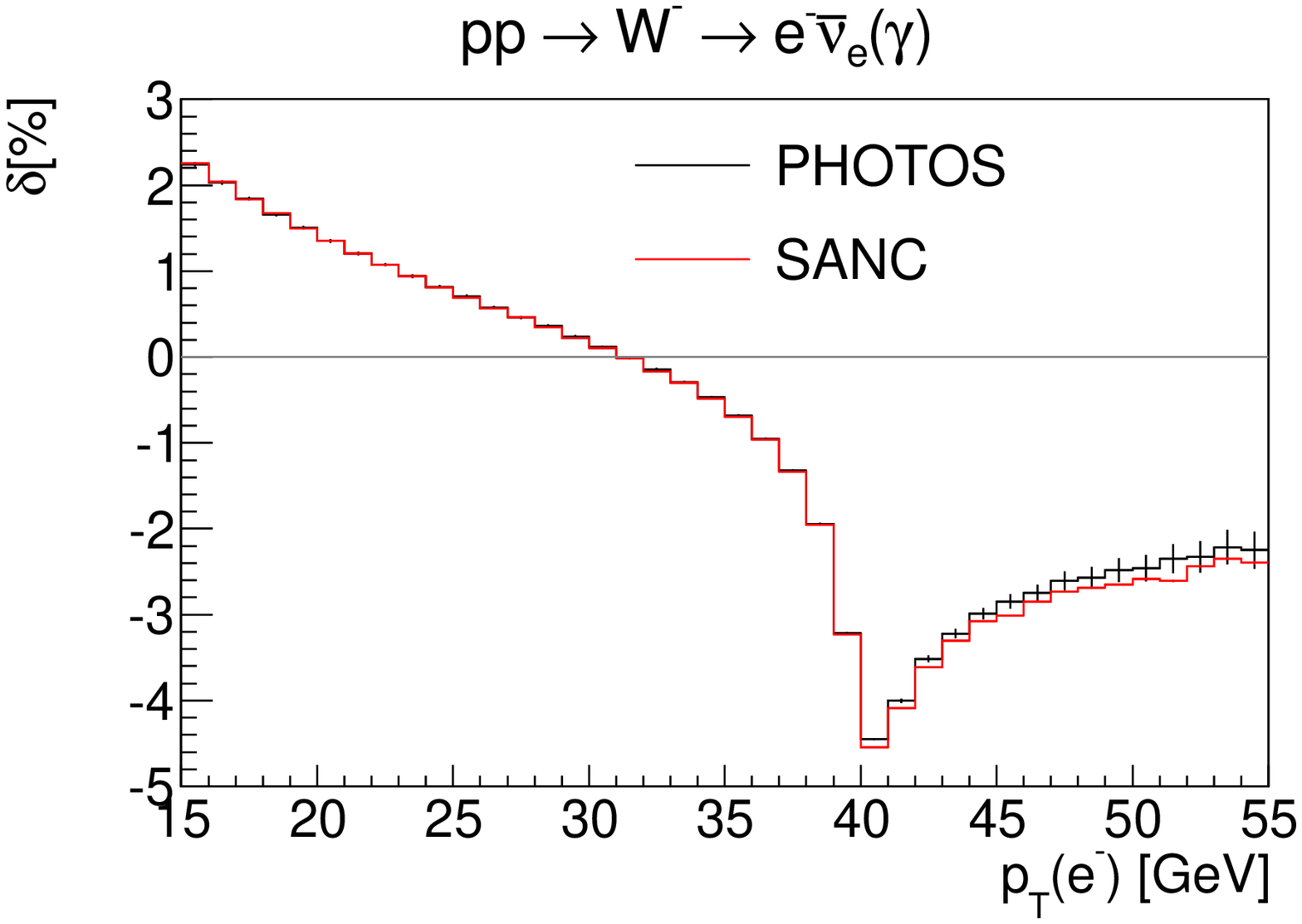}
\includegraphics[width = 0.45\textwidth]{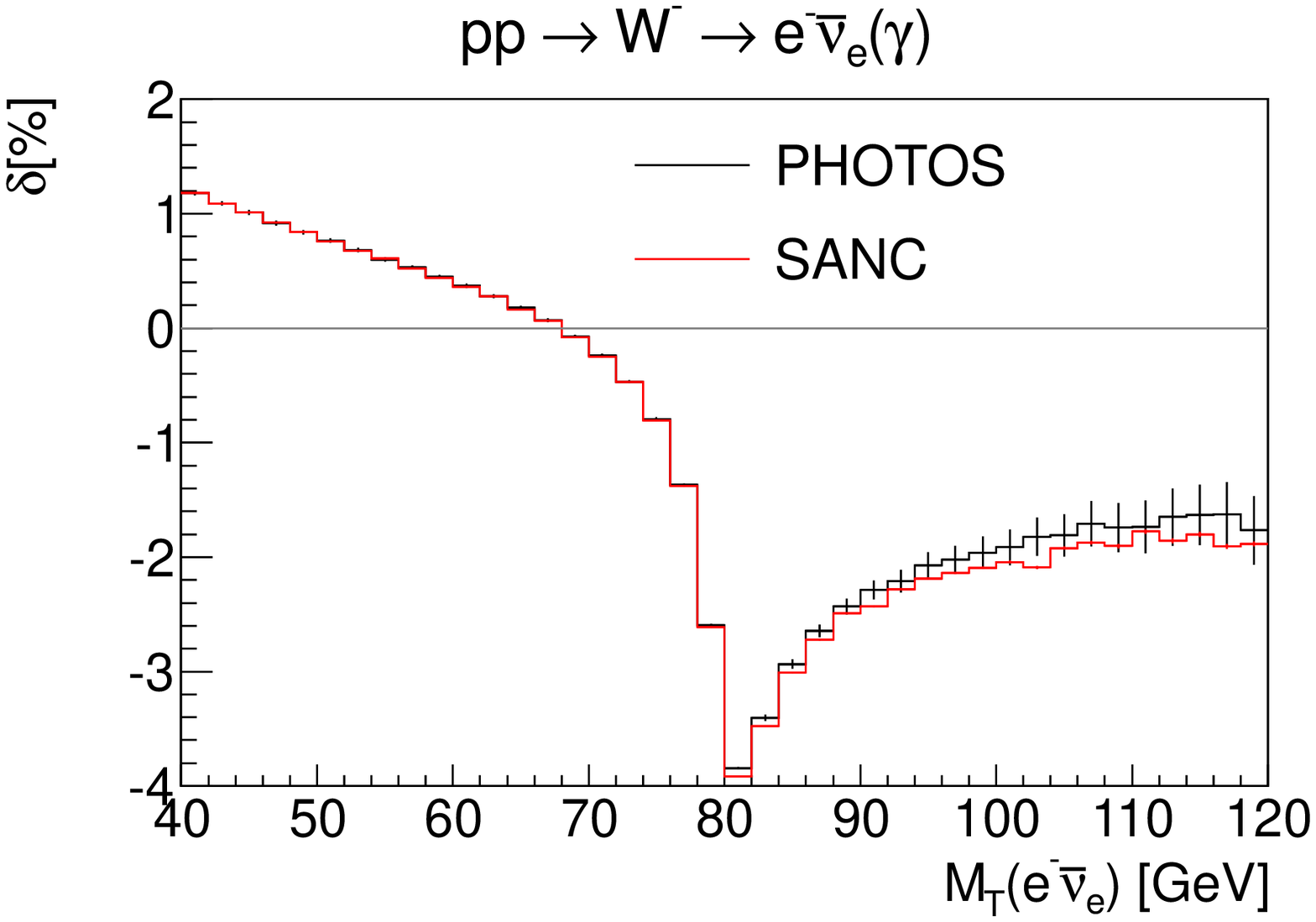}
\end{center}
\caption{\label{fig_sing_cc_calo} Tuned comparison of transverse momentum (left)
and transverse mass (right) distributions in CC DY with calorimetric event selection.}
\end{figure}

Large corrections in the first order make it necessary to look
at higher orders. Fig.~\ref{fig_multi_e} gives an example of comparison
in higher orders. {\tt SANC} computes the second and third order photonic FSR
within the collinear leading logarithmic approximation, thus the results are
sensitive to the choice of the QED factorization scale. {\tt PHOTOS} performs
exponentiation of the first order corrections.
One should also keep in mind that in experimental analysis one typically applies
the so-called calorimetric event selection (where charged leptons are re-combined with
accompanying photons) which reduces the effect and uncertainties
of the FSR considerably, as can be seen from Fig.~\ref{fig_sing_cc_calo}
in comparison with Fig.~\ref{fig_sing_cc_bare}.

\section{Monte Carlo integrator {\tt mcsanc}}

Recently a Monte Carlo integrator (weighted events) based on {\tt SANC} modules~\cite{Andonov:2008ga}
was created~\cite{Bardin:2012jk,Bondarenko:2013nu}. It features:

\begin{itemize}

\item calculations of fully differential cross sections for DY and inclusive cross
sections of higgs-strahlung and single-top production;

\item provides both NLO EW and QCD corrections;

\item support of different EW schemes: $\alpha(0)$, $\alpha(M_Z)$, and $G_\mu$;

\item fixed and running factorization and renormalization scale options;

\item kinematic cuts, recombination are foreseen;

\item parallel calculation on multi-core machines thanks to {\tt Cuba} library\\
\verb"http://www.feynarts.de/cuba/";

\item easy installation and configuration (GNU autotools, {\tt LHAPDF}, input
configurations for physics parameters, cuts, histogramming).

\end{itemize}

The code is suited to be run on multi-core computers.
To get a high precision in differential distributions
it takes many hours. That's why it is important to
take care on stability and rescue. For this reason
integration state files of the {\tt Vegas} algorithm are saved
while running {\tt Cuba}
after every iteration and upon run completion.
The files can be used to increase statistics or
restore from an interrupted run, {\it e.g.} when the batch
time quota has been exceeded.
When the code is run on a multi-core systems, the
calculation is automatically split by the
number of cores with an aid of {\tt {\$}CUBACORES}
environment variable.
The parallelization efficiency is limited
due to inter-process communications:
the optimal number of cores is 8, after
which the run time doesn't reduce and
efficiency (CPU load) remains below 50\%.

The list of processes presently implemented in {\tt mcsanc}
is shown in Table~\ref{tab_mcsanc}. The process id notation is the following: first digit is the sign of EW current,
and the last two digits specify the final particle choice:
\begin{description}
\item [$ 0xx $] means neutral current, $ xx=01(e), 02(\mu), 03(\tau), 04(HZ) $;
\item [$ \pm 1xx $] means charged current, $ xx=01(e)$, $02(\mu)$, $03(\tau)$, $04(HW)$,
$ 05, 06$ (top quark production in $s$ and $t$ channels).
\end{description}

\begin{table}[ht]
\caption{\label{tab_mcsanc}List of processes implemented in {\tt mcsanc}}.
\begin{center}
\begin{tabular}{lcc}
\hline
pid & $ f + f \to $ & {\tt SANC} ref.\\
\hline
001:003 & $ \ell^+ + \ell^- (\ell=e, \mu, \tau) $ &  \cite{Arbuzov:2007db,Andonov:2009nn}
\\
004 & $ Z^0 + H $ & \cite{Bardin:2005dp,Andonov:2008ga} \\
\hline
$ \pm $ 101:103 & $ \ell^\pm +\nu_\ell $ & \cite{Arbuzov:2005dd} \\
$ \pm $ 104 & $ W^\pm +H $ & -\\
\hline
105 & $ t + \bar{b} $ ($s$-channel) & \cite{Bardin:2012jk,Bardin:2012up} \\
106 & $ t + q $ ($t$-channel) & -//- \\
-105 & $ \bar{t} + b $ ($s$-channel) & -//-\\
-106 & $ \bar{t} + q $ ($t$-channel) & -//-\\
\hline
\end{tabular}
\end{center}
\end{table}

\section{Conclusions}

A brief overview of the {\tt SANC} project is presented.
The status of tuned comparisons between the results of {\tt SANC} and other codes for EW corrections
to Drell-Yan production is discussed.

The comparison of {\tt SANC} and {\tt PHOTOS} for single and multiple
photon emission was performed for the neutral and charged current DY processes. The results agree within 0.1\%.

New Monte Carlo integrator {\tt mcsanc-v1.01} is available to download on {\verb"http://sanc.jinr.ru"}.
The tool is aimed for calculation of NLO EW and QCD corrections to Drell-Yan, higgs-strahlung, and single top quark
production processes in $pp$ collisions.

{\bf Acknowledgements.}
This work was supported by RFBR-CERN grant {\verb"12-02-91528-ЦЕРН_а"}
and by Heisenberg-Landau Program grant. AA and RS thank the ACAT organizing committee for partial
covering of expenses during the conference.
AA and AS are grateful for support to the Dynasty foundation.

\end{document}